\documentclass[12pt,preprint]{aastex}

\slugcomment{Accepted for publication in the Astrophysical Journal
Letters}

\shorttitle{Silicate emission in AGN} \shortauthors{Sturm et al.}

\begin{document}

\title{Silicate emissions in active galaxies - From LINERs to QSOs}

\author{
E. Sturm
, M. Schweitzer
, D. Lutz
, A. Contursi
, R. Genzel
, M.D. Lehnert
, and L.J. Tacconi
}
\affil{Max-Planck-Institut
f\"ur extraterrestrische Physik, Postfach 1312, D-85741 Garching,
Germany} \email{sturm@mpe.mpg.de}

\author{S. Veilleux, 
D.S. Rupke 
and D.-C. Kim
} \affil{Department of Astronomy, University of
Maryland, College Park, MD 20742, USA}

\author{A. Sternberg  
 and D. Maoz
}
\affil{School of Physics and Astronomy
Tel Aviv University, Ramat Aviv, Tel Aviv 69978, Israel}

\author{S. Lord 
 and J. Mazzarella
} \affil{IPAC/Caltech, MS 100-22, Pasadena, CA 91125, USA}

\and

\author{D.B. Sanders
} \affil{Institute for Astronomy, University, of Hawaii, 2680
Woodlawn Drive, Honolulu, HI 96822, USA}

\begin{abstract}

We report the first detection of $\sim$10 and $\sim$18\,$\mu$m
silicate dust emissions in a low-luminosity active galactic
nucleus (AGN), obtained in Spitzer-IRS 7-37\,$\mu$m spectroscopy
of the Type~1 LINER galaxy NGC\,3998. Silicate emissions in AGN
have only recently been detected in several quasars. Our detection
counters suggestions that silicate emissions are present only in
the most luminous AGN. The silicate features may be signatures of
a dusty ``obscuring torus'' viewed face-on as postulated for
Type~1 AGN. However, the apparently cool ($\sim$200 K) dust is
inconsistent with theoretical expectations of much hotter torus
walls. Furthermore, not all Type~1 objects are silicate emission
sources. Alternatively, the silicate emission may originate in
dust not directly associated with a torus. We find that the
long-wavelength ($\gtrsim$20\,$\mu$m) tail of the emission in
NGC\,3998 is significantly weaker than in the sample of bright
QSOs recently presented by Hao et al. The 10\,$\mu$m profile in
our NGC\,3998 spectrum is inconsistent with ``standard'' silicate
ISM dust. This may indicate differences in the dust composition,
grain size distribution, or degree of crystallization.  The
differences between NGC\,3998, QSOs, and Galactic templates
suggest that there are significant environmental variations.

\end{abstract}


\section{Introduction}
\label{s:introduction}

In the unifying scheme for active galactic nuclei (AGN) it is
postulated that an ``obscuring torus'' usually surrounds the
accreting massive black hole. Models predict that the spectral
energy distribution (SED) of a torus depends sensitively on its
orientation, geometry, and density distribution (Veilleux 2004,
and references therein). In particular, the tori are predicted to
exhibit prominent silicate dust features in either absorption or
emission, depending on whether an AGN is viewed with the torus
edge-on (Type~2) or face-on (Type~1).

Many Type~2 AGN do indeed exhibit silicate absorption features.
However, until very recently Type~1 objects have not shown any
clear evidence of silicate features, neither in emission nor in
absorption (e.g.~Clavel et al. 2000). In uniform density torus
models with standard dust, suppression of the silicate features
occurs for only a very narrow range of model parameters. To
account for the non-detections clumpy torus models or tapered
disks have been invoked (e.g., Nenkova et al. 2002, Efstathiou \&
Rowan-Robinson 1995), and larger grain sizes have been postulated
(e.g., Laor \& Draine 1993; Maiolino et al. 2001a, 2001b).

The Spitzer Space Telescope with its good wavelength coverage and
excellent sensitivity allows a more detailed re-investigation of
this problem. Siebenmorgen et al. (2005) and Hao et al. (2005)
have reported the first detections of prominent silicate emission
features in mid-infrared Spitzer-IRS spectra of several luminous
quasars. Here, we report the first such detection in a low
luminosity AGN, the LINER galaxy NGC\,3998. We also comment on
detections in more luminous AGN in our sample, with luminosities
up to 10$^4$ greater than in NGC\,3998.


NGC\,3998 is an S0 galaxy at a distance of 14.1 Mpc (Tonry et al.
2001). The optical line ratios are typical of Low-Ionization
Nuclear Emission-Line Regions (LINERs). Because of a broad
H$\alpha$ component (Ho et al. 1997) NGC\,3998 has been classified
as a Type~1 object. The 2-10 keV X-ray luminosity of NGC\,3998 is
~3x10$^{41}$ erg/s (Ptak et al. 2004). A strong UV point source
was detected by HST (Fabbiano et al. 1994).  The UV source is
variable indicating AGN accretion, as opposed to a stellar, energy
source (Maoz et al.~2005).

\section{Observations and Data Processing}
\label{s:obs}

Our data were obtained with the Spitzer IRS (Houck et al. 2004;
Werner et al. 2004) as part of our GO Cycle-1 project on LINERs.
For comparison we also consider PG QSOs which are part of our medium
size GO Cycle-1 program on Quasar and ULIRG evolution (QUEST; PI:
S. Veilleux). The objects were observed in three low and high
resolution IRS modules: short low (S0), short high (S1) and long
high (S3). Our data reduction started with the two-dimensional BCD
products from the Spitzer pipeline (S11). We used our own IDL
routines for de-glitching and sky subtraction, and SMART (Higdon
et al. 2004) for extraction of the final spectrum.

\section{Results and Discussion}
\label{s:results}

\begin{figure}[t]
\plotone{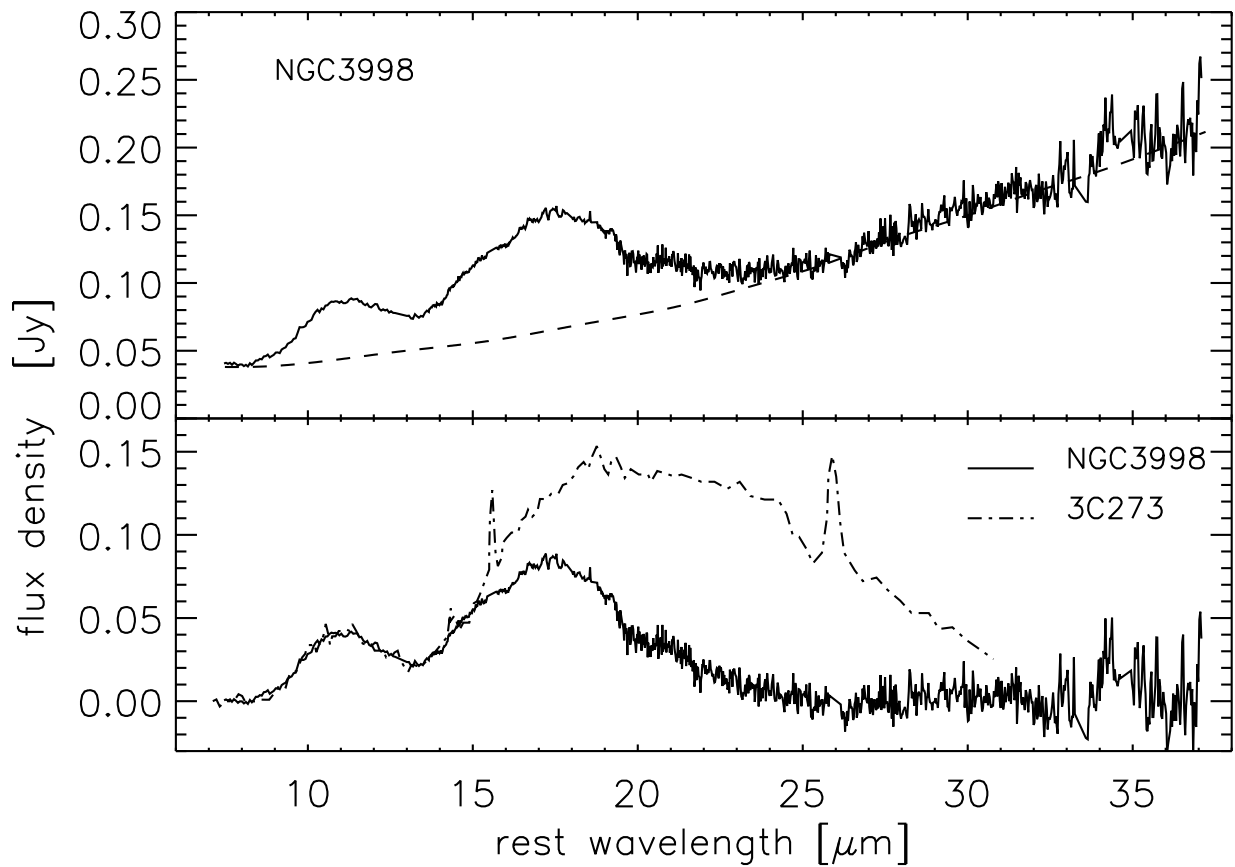} \caption{{\it Top}: Combined low and high
resolution IRS rest frame spectra of NGC\,3998. Narrow emission
lines have been omitted. The continuum used for the bottom part of
this figure is indicated. {\it Bottom}: The continuum subtracted
spectrum of NGC\,3998 compared to the PG QSO 3C273 (from Hao et
al. 2005). No scaling has been applied.} \label{F:n3998}
\end{figure}

The SED of NGC\,3998 is characterized by a well-detected warm
continuum with clear broad emission features in the 9 - 13\,$\mu$m
and 15 - 20\,$\mu$m range (Fig. \ref{F:n3998}). PAH dust emission
features are absent. The spectrum also exhibits a number of ionic
emission lines (omitted in Fig. 1), that will be analyzed
elsewhere.



Siebenmorgen et al. (2005) have suggested that the failure to
detect silicate emission in AGN in earlier missions could be
explained as luminosity dependence, because the only detections so
far have been in highly luminous (but apparently faint) quasars.
Our detection of a prominent silicate feature in a LINER with an
AGN luminosity 4 to 5 orders of magnitude below those of quasars
is clear evidence against this hypothesis. Silicate emission can
be equally strong in low luminosity AGN (see the comparison in
Fig.~\ref{F:n3998}). The AGN in which silicate emissions have been
reported to date are listed in Table~\ref{tab:AGNcollection}. This
emission is present for a large range of intrinsic 2-10 keV X-ray
luminosity. Our QUEST and LINER programs are beginning to provide
additional observations. Although only a fraction of these
observations have been taken or analyzed so far, these surveys
provide an early indication of the prevalence of silicate
emissions in AGN. They appear in a majority of our optically
selected QSOs (with 2-10 keV X-ray luminosities ranging from
$\approx$10$^{43.3}$ to 10$^{44.3}$ erg/s) but in only a small
fraction of LINERs (1 out of 10). In many LINERs (but not in
NGC\,3998) and Seyferts the analysis is complicated by confusion
with strong PAH emission, requiring a careful decomposition of the
spectra into the various components. This will make determining
the prevalence and characteristics of silicate emission
challenging.


In Fig.~\ref{F:n3998} we compare NGC\,3998 to one of the QSOs
(3C273) in the Hao et al. (2005) sample, using their continuum
subtracted spectrum. For NGC\,3998 we subtracted the continuum by
fitting a spline (see Fig.~\ref{F:n3998}) through the continuum
points in the 7-8\,$\mu$m and 25-35\,$\mu$m ranges. Without any
further scaling the two spectra are virtually identical in the
10\,$\mu$m feature, but at wavelengths $\gtrsim$20\,$\mu$m they
deviate significantly from each other. The other QSO spectra also
match the shape of the 10\,$\mu$m feature but the peak wavelength
varies between ~10 and ~11.5\,$\mu$m.  All are different from NGC
3998 at longer wavelengths.

\begin{figure}[t]
\epsscale{1.0} \plotone{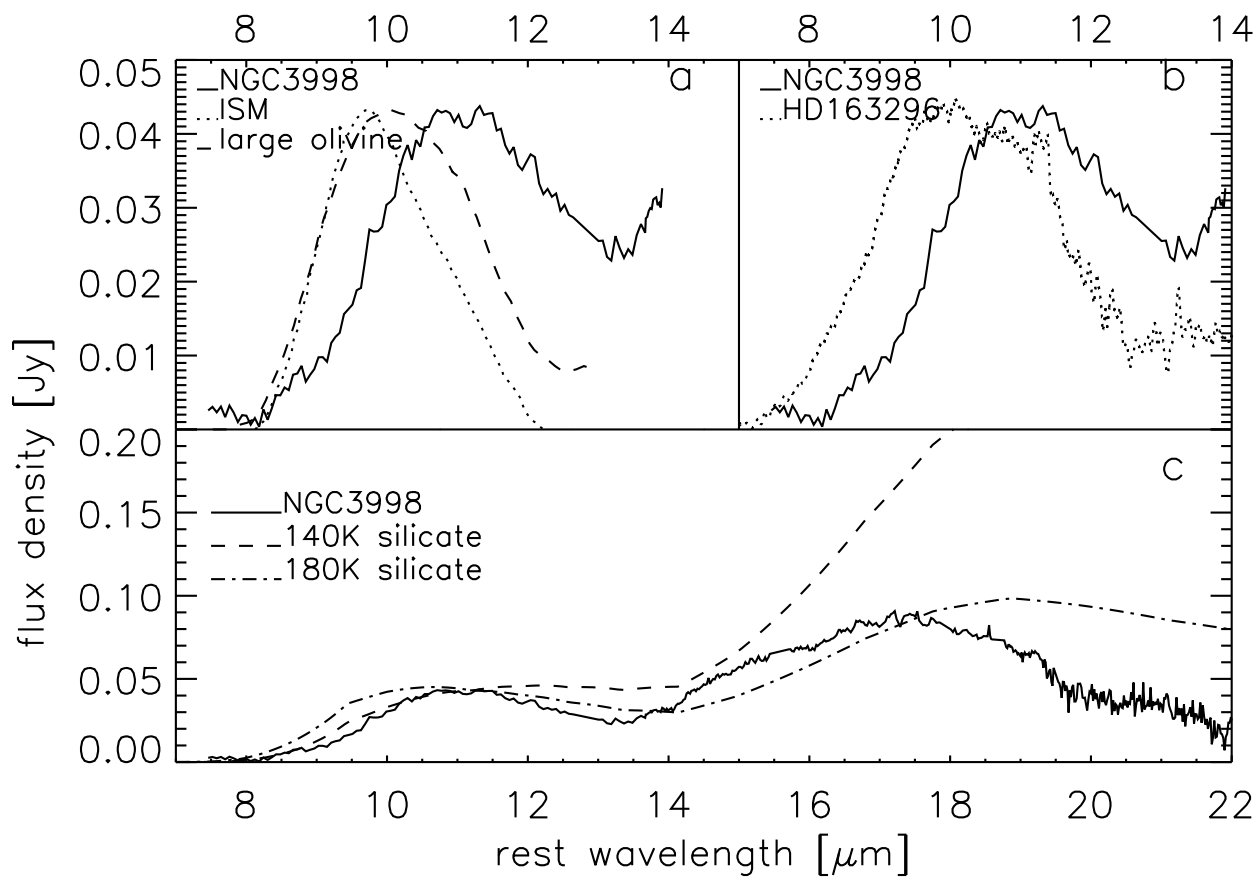} \caption{(a): Comparison of the
NGC\,3998 10\,$\mu$m feature to the standard ISM emission of
amorphous silicates (dotted, Kemper et al. 2004), and to olivines
of larger grain size (dashed, van Boekel et al. 2005). (b):
Comparison to the Herbig Be star HD163296 (dotted, Meeus et al.
2001). (c): Dashed (dash-dotted): the product of a 140K (180K)
blackbody and the ISM silicate opacities, produced as in Hao et
al. (2005).} \label{F:comp_templates}
\end{figure}

We now compare the 10\,$\mu$m profile in NGC\,3998 with feature
``templates'' from various astronomical environments. Figure
\ref{F:comp_templates}a shows a comparison to normal ISM silicate
emission (Kemper et al.~2004). Clearly, the normal ISM profile is
not a good fit. The ISM silicate profile peaks at
$\sim$9.8\,$\mu$m, and it is narrower than in NGC\,3998.
Siebenmorgen et al. and Hao et al. try to fit the broadened and
shifted profiles by folding ISM dust opacities with steeply rising
(cold) Planck functions. However, in NGC\,3998 this cannot be the
full explanation (Fig.~\ref{F:comp_templates}c). While the
relative intensities of the 10/18 peaks in NGC\,3998 can be
reproduced by such a fit with a temperature of $\sim$180\,K, the
emission in the blue wing of the 10\,$\mu$m feature and longwards
of $\sim$20$\mu$m is significantly over-predicted.

Fig.~\ref{F:comp_templates}a also shows a template spectrum for
enhanced grain sizes (van Boekel et al.~2005). An increase in the
grain size clearly shifts and broadens the profile. Van Boekel et
al. have used such templates to fit silicate emission profiles of
young stellar objects (Herbig Be stars). They found that none of
the stellar sources consist of fully pristine dust comparable to
that found in the ISM, and that larger grain sizes and/or
crystallization due to dust processing are necessary to fit the
spectra. In Fig.~\ref{F:comp_templates}b we show a comparison of
NGC\,3998 with the Herbig Be star HD163296 (Meeus et al.~2001).
The shapes of the two 10\,$\mu$m {\it profiles} match quite well;
the emission {\it peak} in NGC\,3998 is shifted towards longer
wavelengths, however.

While an increased average grain size of amorphous silicates
(olivine, pyroxene) can move the feature as a whole, an increased
admixture of crystalline silicates (fosterites, enstatites) would
result in an additional, relatively sharp peak at 11.3\,$\mu$m and
further features at longer wavelengths (e.g. Meeus et al. 2001).
Within the S/N limits we do not see clear evidence for such peaks
in NGC\,3998. We note, however, that other species of crystalline
silicates, like clino-pyroxenes, produce weaker discrete features
and would also be consistent with the faster decline at
$\sim$18\,$\mu$m (Wooden et al. 1999).

The feature profile may also be sensitive to the chemical
composition. For instance, ISO spectra of carbon-rich environments
around red giants revealed strong emission of SiC around
11.2\,$\mu$m (e.g. Aoki et al. 1999), which looks very reminiscent
of the features studied here. Jaffe et al. (2004) reported a
similar blue wing deviation and peak shift in the {\it absorbed}
silicate profile in NGC1068. This indicates that the observed
shift of silicate emission may not be purely a temperature effect.


We conclude that, due to dust processing, the dust giving rise to
the emission features in these AGN may deviate from pristine
galactic ISM dust. Evidence that dust in the circum-nuclear region
of AGN has different properties (like larger grain size) than in
the Galactic diffuse ISM has been reported earlier (Maiolino et
al. 2001a, 2001b). Our conclusion does not depend on the
simplified assumption of a single temperature that we made in the
fit described above (ISM dust opacities x Planck function). We
have repeated the fit using a mixture of various Planck functions
with cooler and warmer temperatures. Applying a mixture of
temperatures rather than a single temperature worsens the fit: the
silicon features are broadened even more, such that the mismatch
in the blue wing of the 10\,$\mu$m feature and the red wing of the
18\,$\mu$m feature increases.

The presence of the 10\,$\mu$m silicate emission feature over a
broad range of AGN luminosities could be taken as evidence for the
obscuring torus. However, it is not clear yet whether the silicate
emission arises in fact from a face-on torus of the standard
unified AGN scenario. Apparently, some AGN have a significant
region of optically thin, relatively cool ($\sim$200\,K) dust
emission with comparable properties in various AGN types. This
temperature, though, seems too cold to be explained by a hot inner
torus wall, which should emit closer to the sublimation
temperature of silicates. Depending on the size and composition of
the grains, this is between $\approx$800 and 1500\,K (Kimura et
al. 2002). A lower temperature might be the result of a
stratification into the torus wall, with cooler dust from deeper
layers contributing to the inferred average value. However, as
described above, mixing emission components of different
temperature makes the fit to the observed feature profiles worse.
This preliminary approach needs to be improved by real
(3-dimensional) radiative transfer models of the AGN emission,
before final conclusions can be reached. The case of NGC\,3998
with its low AGN luminosity and different SED shape at longer
wavelengths compared to QSOs places important new constraints on
such modeling.

More indication that the silicate emission may not originate in a
(standard) torus is provided by the fact that other Type~1 AGN
(e.g. some QUEST QSOs and some QSOs of the Hao et al. sample) do
{\it not} show silicates in emission.
Furthermore, obscuring tori may not be present in LINERs at all,
given that variable central UV point sources are observed in both
Type~1 and Type~2 LINERs (Maoz et al.~2005).
In addition, observations of nearby Seyfert galaxies have revealed
extended dust emission that is roughly co-spatial with the NLR
(e.g.~Cameron~1993; Bock et al.~1998; Tomono et al.~2001; Radomski
et al.~2003; Packham et al.~2005). The role and influence of this
extended component is currently poorly constrained but may be
significant, given the absence of a strong Type~1 vs. Type~2
anisotropy in total AGN infrared continua (Lutz et al.~2004). We
note that dust temperatures of the order 200\,K are fully
consistent with observed color temperatures of extended NLR dust
(Radomski et al.~2003, Packham et al.~2005). Future results from
Spitzer observations of both Type~1 and Type~2 AGN over the entire
range of intrinsic luminosity will certainly shed more light on
this issue.

\acknowledgments

This work is based on observations made with the Spitzer Space
Telescope, which is operated by the Jet Propulsion Laboratory,
California Institute of Technology under NASA contract 1407.
Support for this work was provided by NASA through Contracts
\#1263752 and \#1267948 issued by JPL/Caltech (SV, DSR, DCK). AS
thanks the Israel Science Foundation for support.

\clearpage

\begin{table}[t]
\caption{AGN with silicate
emission.\label{tab:AGNcollection}}
\begin{tabular}{llll}
\tableline\tableline
Object          & Type      & z         & L$_{2-10keV}$  \\
                &           &           & [10$^{43}$erg/s]  \\
\tableline
NGC\,3998\tablenotemark{a}         & Liner 1 & 0.0035          &  0.03      \\
PG0804+761\tablenotemark{b}      & QSO       & 0.0999    &  29         \\
PG1211+143\tablenotemark{b}      & QSO       & 0.0808    &  5         \\
PG1351+640\tablenotemark{b}      & QSO       & 0.0881    &           \\
IZw1 (PG0050+124)\tablenotemark{b}   &RL QSO & 0.0611    &  7.8         \\
3C273 (PG1226+023)\tablenotemark{b}  &RL QSO & 0.158     &  510         \\
3C249.1 (PG1100+772)\tablenotemark{c}&RL QSO & 0.312     & 110          \\
3C351 (PG1704+608)\tablenotemark{c}  &RL QSO & 0.372     &           \\
\tableline
\end{tabular}
\tablenotetext{a}{This work} \tablenotetext{b}{Hao et al. (2005)}
\tablenotetext{c}{Siebenmorgen et al. (2005) }
\end{table}

\clearpage

\clearpage

%

\end{document}